\documentclass[12pt]{article}
\usepackage{epsfig}
\usepackage{a4}
\usepackage{latexsym}
\usepackage{cite}

\usepackage{pslatex}
\usepackage[latin1]{inputenc}
\usepackage[T1]{fontenc}

\usepackage{color,colordvi}
\def\colour4colour#1{\Blue{#1}}

\newcommand{\gsim}{\raisebox{-0.07cm}{$\:\:\stackrel{>}{{\scriptstyle
 \sim}}\:\: $} }

\newcommand{\beq}{\begin{equation}}
\newcommand{\eeq}{\end{equation}}
\newcommand{\bea}{\begin{eqnarray}}
\newcommand{\eea}{\end{eqnarray}}
\newcommand{\nn}{\nonumber}

\newcommand{\als}{\alpha_{\rm s}}
\newcommand{\Qs}{Q^{\:\! 2}}
\newcommand{\as}{a_{\rm s}}
\newcommand{\ep}{\epsilon}
\newcommand{\ra}{\rightarrow}

\newcommand{\hspn}{{\hspace{-3mm}}}

\begin{document}
\setlength{\parskip}{0.2cm}
\setlength{\baselineskip}{0.535cm}

\def\plus{{\!+\!}}
\def\minus{{\!-\!}}
\def\z#1{{\zeta_{#1}}}
\def\zs2{{\zeta_{2}^{\,2}}}
\def\ca{{C^{}_A}}
\def\cas{{C^{\, 2}_A}}
\def\cat{{C^{\, 3}_A}}
\def\cf{{C^{}_F}}
\def\cfs{{C^{\, 2}_F}}
\def\cft{{C^{\, 3}_F}}
\def\nf{{n^{}_{\! f}}}
\def\nfs{{n^{\,2}_{\! f}}}
\def\nft{{n^{\,3}_{\! f}}}

\def\dabc2n{{{d^{abc}d_{abc}}\over{n_c}}}

\def\pqq(#1){p_{\rm{qq}}(#1)}
\def\pqg(#1){p_{\rm{qg}}(#1)}
\def\pgq(#1){p_{\rm{gq}}(#1)}
\def\pgg(#1){p_{\rm{gg}}(#1)}
\def\H(#1){{\rm{H}}_{#1}}
\def\Hh(#1,#2){{\rm{H}}_{#1,#2}}
\def\Hhh(#1,#2,#3){{\rm{H}}_{#1,#2,#3}}
\def\Hhhh(#1,#2,#3,#4){{\rm{H}}_{#1,#2,#3,#4}}

\begin{titlepage}
\noindent
DESY 07-155  \hfill {\tt arXiv:0709.3899 [hep-ph]}\\
SFB/CPP-07-53 \\
LTH 758 \\[1mm]
September 2007 \\
\vspace{1.4cm}
\begin{center}
\Large {\bf On Third-Order Timelike Splitting Functions} \\[2mm]
\Large {\bf and Top-Mediated Higgs Decay into Hadrons} \\
\vspace{2.4cm}
\large
S. Moch$^{\, a}$ and A. Vogt$^{\, b}$ \\
\vspace{1.5cm}
\normalsize
{\it $^a$Deutsches Elektronensynchrotron DESY \\
\vspace{0.1cm}
Platanenallee 6, D--15738 Zeuthen, Germany}\\
\vspace{0.5cm}
{\it $^b$Department of Mathematical Sciences, University of Liverpool \\
\vspace{0.1cm}
Liverpool L69 3BX, United Kingdom}\\
\vfill
\large {\bf Abstract}
\vspace{-0.2cm}
\end{center}
We employ relations between spacelike and timelike deep-inelastic processes 
in perturbative QCD to calculate the next-to-next-to-leading order (NNLO) 
contributions to the timelike quark-quark and gluon-gluon splitting functions 
for the evolution of flavour-singlet fragmentation distributions. We briefly 
address the end-point behaviour and the numerical size of these third-order 
corrections, and write down the second moments of all four timelike splitting 
functions. In the same manner we re-derive the NNLO result for the Higgs-boson 
decay rate into hadrons in the limit of a heavy top quark and five massless 
flavours, and confirm the recent N$^3$LO computation of this quantity.
\vspace{0.5cm}
\end{titlepage}
\noindent
In this letter we present new results on the scale dependence (evolution) of 
the parton fragmentation distributions $\,D_{\! f}^{\,h}(x,\Qs)$. Here $x$ 
denotes the fraction of the momentum of the final-state parton $f$ carried by 
the outgoing hadron $h$, and $\,\Qs$ is a timelike hard scale, such as the 
squared four-momentum of the gauge boson in $\, e^+\,e^- \ra\, \gamma\,,\:Z 
\,\ra\, h + X$. This scale dependence is given by
\beq
\label{eq:Devol}
  {d \over d \ln \Qs} \; D_{i}^{\,h} (x,\Qs) \:\: = \:\:
  \int_x^1 {dz \over z} \; P^{\,T}_{ji} \left( z,\als (\Qs) \right)
  \:  D_{j}^{\,h} \Big( {x \over z},\, \Qs \Big) 
\eeq
where the summation over $\,j \, =\, q,\:\bar{q},\:g\,$ is understood. 
The timelike splitting functions $\,P^{\,T}_{ji}\,$ admit an expansion in
powers of the strong coupling $\als$, 
\beq
\label{eq:PTexp}
  P^{\,T}_{ji} \left( x,\als (\Qs) \right) \:\: = \:\:
  \as \, P_{ji}^{(0)\,T}(x) \: + \: \as^{\:\!2} \, P_{ji}^{(1)\,T}(x)
  \: +\: \as^{\:\!3} P_{ji}^{(2)\,T}(x) \: +\: \ldots \:\: ,
\eeq
where we normalize the expansion parameter as $\,\as \equiv \als(\Qs)/ (4\pi)$.
The leading-order (LO) terms in Eq.~(\ref{eq:PTexp}) are identical to the 
spacelike case of the initial-state parton distributions, a fact often 
referred to as the Gribov-Lipatov relation \cite{Gribov:1972ri}. 
Also the next-to-leading order (NLO) contributions $\,P_{ji}^{(1)\,T}\!(x)$ 
have been known for more than 25 years. These quantities are related to their 
spacelike counterparts by a suitable analytic continuation 
\cite{Curci:1980uw,Furmanski:1980cm,Floratos:1981hs}, see also Ref.~\cite
{Stratmann:1996hn}.

In a previous publication \cite{Mitov:2006ic} we have calculated the next-to-%
next-to-leading order (NNLO) splitting functions $\,P_{\rm ns}^{(2)\,T}\!(x)$ 
for the non-singlet combinations of quark fragmentation distributions. This
calculation was based on an analytic continuation of the corresponding 
unrenormalized partonic structure function in deep-inelastic scattering (DIS)
\cite{Moch:2004pa}, performed after subtracting contributions due to the quark
form factor \cite{Moch:2005id} and backed up by another relation between the 
spacelike and timelike cases conjectured in Ref.~\cite{Dokshitzer:2005bf}.
We now address the flavour-singlet timelike evolution
\beq
\label{eq:Dsgevol}
  \frac{d}{d \ln \Qs}
  \left( \begin{array}{c} \!D^{}_{\!S}\! \\ \!D_{\!g}\! \end{array} \right)
  \: = \: \left( \begin{array}{cc} 
         P_{\rm qq}^{\,T} & P_{\rm gq}^{\,T} \\[2mm] 
         P_{\rm qg}^{\,T} & P_{\rm gg}^{\,T}
   \end{array} \right) \otimes
  \left( \begin{array}{c} \!D^{}_{\!S}\! \\ \!D_{\!g}\! \end{array} \right)
  \quad \mbox{with} \qquad
  D^{}_{\!S} \; \equiv \; \sum_{r=1}^{\nf} ( D_{q_r^{}} + D_{\bar{q}_r^{}} ) 
  \:\: .
\eeq
Here $\otimes$ abbreviates the Mellin convolution written out in Eq.~(\ref
{eq:Devol}), and $\nf$ stands for the number of effectively massless quark
flavours. Specifically, we extend the approach of Ref.~\cite{Mitov:2006ic} to 
derive the NNLO diagonal entries $\,P_{\rm qq}^{(2)\,T}\!(x)$ and 
$\,P_{\rm gg}^{(2)\,T}\!(x)$ in Eq.~(\ref{eq:Dsgevol}). The relation of this 
calculation to Higgs-boson decay will be addressed below. 

For brevity focusing on the gluonic case, our calculation starts form the 
unrenormalized structure function $F_{\phi,g}^{\,\rm b}$ for DIS by the 
exchange of a scalar $\phi$ coupling (like the Higgs-boson in the heavy-top 
limit) directly only to gluons via $\phi\,G_{\mu\nu}^{\,a} G_a^{\:\mu\nu}\!$, 
where $G^{\,a}_{\mu\nu}$ denotes the gluon field strength tensor. 
This quantity has been computed to three loops for the determination of the 
spacelike NNLO quark-gluon and gluon-gluon splitting functions in 
Ref.~\cite{Vogt:2004mw}. In dimensional regularization with $\,D = 4 - 2\ep\,$ 
its perturbative expansion in terms of the bare reduced coupling 
$\as^{\,\rm b}$ can be written as
\beq
\label{eq:Fphib}
  F_{\phi,g}^{\,\rm b} (\as^{\,\rm b}, Q^2) \; = \;
  \delta(1-x) \:  + \: \sum_{l=1}^{\infty} \: ( \as^{\,\rm b\:\!} )^n
  \left( Q^2 \over \mu^2 \right)^{-n\ep} F_{\phi,g}^{\,{\rm b} (n)} \:\: .
\eeq
The $n$-th order terms $F_{\phi,g}^{\,{\rm b} (n)}$ are then iteratively 
decomposed into contributions arising from the analogous expansion coefficients 
${\cal F}^{}_{\!m \leq n}$ of the $\,\phi gg\,$ form factor \cite{Moch:2005tm} 
and remaining `real' parts ${\cal R}^{}_{\:n}$, \pagebreak
\bea
\label{eq:FphiFR}
F_{\phi,g}^{\,\rm b (1)}
     &\!=\!& 2 {\cal F}_1\,\delta(1-x) + {\cal R}_{\,1} \nn \\[0.5mm]
F_{\phi,g}^{\,\rm b (2)}
     &\!=\!& 2 {\cal F}_2\, \delta(1-x)
           + \left({\cal F}_1\right)^2 \delta(1-x)
           + 2 {\cal F}_1 {\cal R}_{\,1} + {\cal R}_{\,2} \nn \\[0.5mm]
F_{\phi,g}^{\,\rm b (3)}
     &\!=\!& 2 {\cal F}_3\, \delta(1-x)
           + 2 {\cal F}_1 {\cal F}_2\, \delta(1-x)
           + (\, 2 {\cal F}_2 + \left({\cal F}_1\right)^2 \,)\, {\cal R}_{\,1}
           + 2 {\cal F}_1 {\cal R}_{\,2} + {\cal R}_{\,3}
\:\: .
\eea
Note that the functions ${\cal R}_{\:m \geq 2}$ do not only collect tree-level 
amplitudes but also combinations of real-emission and virtual corrections.
As in Ref.~\cite{Mitov:2006ic}, this will lead to a problem in the third-order
analytic continuation. For $P_{\rm gg}^{\,(2)T}\!$, however, this problem can 
be fixed afterwards in complete analogy to the previous non-singlet quark case 
(see below), thus we can ignore it for the moment.

The analytic continuation of the form factor to the time-like case is known.
The $x$-dependent functions ${\cal R}_{\:n}$ are continued from $x$ to $1/x$
\cite{Curci:1980uw,Furmanski:1980cm,Floratos:1981hs,Stratmann:1996hn}, 
taking into account the (complex) continuation of $q^2$ (see Eq.~(4.1) of 
Ref.~\cite{Moch:2005id}$\,$) and the additional prefactor $x^{1-2\ep}$ 
originating from the phase space of the detected parton in the timelike case
\cite{Rijken:1996ns}. We have performed this continuation using routines for 
the harmonic polylogarithms (HPLs) \cite{Remiddi:1999ew} implemented in 
{\sc Form} \cite{Vermaseren:2000nd}.
The only subtle point in the analytic continuations is the treatment of 
logarithmic singularities for $x \ra 1$ starting with
$$
\label{eq:l1xAC}
  \ln (1-x) \:\:\ra\:\: \ln (1-x) - \ln x + i\,\pi \:\: .
$$

\vspace{-2mm}
After these analytic continuations the one-parton inclusive fragmentation 
function in $\phi$-decay is re-assembled order by order analogous to Eq.~(\ref
{eq:FphiFR}), keeping the real parts of the continued ${\cal R}_{\:n}$ only. 
Then the renormalization of the operator $G_{\mu\nu}^{\,a} G_a^{\:\mu\nu}$ and 
the strong coupling constant is performed. Finally the timelike splitting 
functions (and coefficient functions) can extracted iteratively from the mass
factorization relations
\bea
\label{eq:FphiT1}
 F_{\phi,g}^{\,(1)T} &\!=\!\!& \mbox{}
   - {1 \over \ep}\, P_{\rm gg}^{\,(0)} \: +\: c^{\,(1)T}_{\phi,\rm g} 
   \: +\: \ep\, a^{\,(1)T}_{\phi,\rm g} 
   \: +\: \ep^2\, b^{\,(1)T}_{\phi,\rm g} \; +\; \ldots
\\[1mm]
\label{eq:FphiT2}
 F_{\phi,g}^{\,(2)T} &\!=\!\!& \;
   {1 \over 2\ep^2}\, 
     \bigg\{ \!\! \left( P_{\rm gi}^{(0)} + \beta_0 \delta_{\rm gi} \right) 
      P_{\rm ig}^{(0)} \bigg\}
   \: - \: {1 \over 2\ep}\, 
     \bigg\{ P_{\rm gg}^{\,(1)T} + 2\, P_{\rm gi}^{(0)} c^{\,(1)T}_{\phi,\rm i}
     \bigg\} 
   \: + \:  c^{\,(2)T}_{\phi,\rm g} - P_{\rm gi}^{(0)} a^{\,(1)T}_{\phi,\rm i}
\nn \\ & & \mbox{}
   \: + \: \ep\, 
     \bigg\{ a^{\,(2)T}_{\phi,\rm g} - P_{\rm gi}^{(0)} b^{\,(1)T}_{\phi,\rm i}
     \bigg\} \; + \; \ldots 
\\[1mm]
\label{eq:FphiT3}
 F_{\phi,g}^{\,(3)T} &\!=\!\!& \mbox{}
   - \: {1 \over 6\ep^3}\, 
     \bigg\{ P_{\rm gi}^{(0)} P_{\rm ij}^{(0)} P_{\rm jg}^{(0)}
       + 3 \beta_0\, P_{\rm gi}^{(0)} P_{\rm ig}^{(0)} 
       + 2\, \beta_0^{\,2}\, P_{\rm gg}^{(0)} 
     \bigg\}
\nn \\ & & \mbox{}
   + \: {1 \over 6\ep^2}\, 
     \bigg\{ 2 P_{\rm gi}^{(0)} P_{\rm ig}^{(1)T} 
       \! + P_{\rm gi}^{(1)T} \! P_{\rm ig}^{(0)} 
       \! + 2\beta_0\, P_{\rm gg}^{(1)T} + 2\beta_{1\,} P_{\rm gg}^{(0)}
       \! + 3 P_{\rm gi}^{(0)} \!\left( P_{\rm ij}^{(0)} + \beta_0 
         \delta_{\rm ij} \right) c^{\,(1)T}_{\phi,\,\rm j} 
     \bigg\}
\nn \\ & & \mbox{}
   - \: {1 \over 6\ep}\,
     \bigg\{ 2\, P_{\rm gg}^{(2)T} 
       + 3\, P_{\rm gi}^{(1)T} \! c^{\,(1)T}_{\phi,\rm i}
       + 6\, P_{\rm gi}^{(0)} c^{\,(2)T}_{\phi,\rm i}
       - 3 P_{\rm gi}^{(0)} \left( P_{\rm ij}^{(0)} + \beta_0 \delta_{\rm ij}
         \right) a^{\,(1)T}_{\phi,\,\rm j}
     \bigg\}
\nn \\ & & \mbox{}
   + \: c^{\,(3)T}_{\phi,\rm g} 
   - {1 \over 2}\, P_{\rm gi}^{(1)T} \! a^{\,(1)T}_{\phi,\rm i} 
   - P_{\rm gi}^{(0)} a^{\,(2)T}_{\phi,\rm i}
   + {1 \over 2}\, P_{\rm gi}^{(0)} \left( P_{\rm ij}^{(0)} + 
     \beta_0 \delta_{\rm ij} \right) b^{\,(1)T}_{\phi,\,\rm j}
   \; + \; \ldots 
\;\; .
\eea
Here all products of $x$-dependent (generalized) functions are to be read
as Mellin convolutions or as products in Mellin-$N$ space, employing routines 
for harmonic sums and their inverse Mellin transform back to $x$-space 
\cite{Vermaseren:1998uu,Vermaseren:2000nd}.
Obviously the determination of $\,P_{\rm gg}^{(2)T}$ from Eq.~(\ref{eq:FphiT3})
requires the `off-diagonal' two-loop coefficient function 
$c^{\,(2)T}_{\phi,\rm q}$.  This quantity, $a^{\,(2)T}_{\phi,\rm q}$ and the
corresponding first-order functions (for all these our normalization differs 
from the standard convention by a factor of two) can be calculated via a direct
analytic continuation of $F_{\phi,q}^{\,\rm b (1,2)}$.
We have checked this fact by comparing the corresponding gluonic results for
the photon-exchange case to an explicit two-loop calculation to order $\ep$
\cite{Mitov:2006wy}. At the third order in $\als$, however, this direct 
continuation fails to correctly reproduce the $\pi^2$ contributions already at 
order $\ep^{-3}$, thus we cannot derive the off-diagonal timelike NNLO 
splitting functions in this simple manner.  

We are now ready to present our results for the diagonal timelike splitting
functions. For completeness we start at NLO where we, of course, reproduce
the results of Ref.~\cite{Furmanski:1980cm}. Adopting the notations of Ref.~%
\cite{Remiddi:1999ew} for the HPLs, the timelike -- spacelike differences can
be written as
\bea
 \lefteqn{ \delta\, P^{\,(1)}_{\rm ps}(x) \;\; \equiv \;\;
           P^{\,(1)T}_{\rm ps}(x) -  P^{\,(1)S}_{\rm ps}(x) \;\; = \;\; }
  \nn\\&& \mbox{}
         8\, \*  \colour4colour{\cf \* \nf }  \*  \Big(
          - 20/9\: \* x^{-1} -3 - x + 56/9\: \* x^2
          - ( 3 + 7\, \* x + 8/3\: \* x^2 )\,\* \H(0)
          + 2 \* (1+x)\, \* \Hh(0,0)
          \Big)
\:\: ,\label{eq:dPps1}
\\[1mm]
 \lefteqn{ \delta\, P^{\,(1)}_{\rm gg}(x) \;\; \equiv \;\;
           P^{\,(1)T}_{\rm gg}(x) -  P^{\,(1)S}_{\rm gg}(x) \;\; = \;\; }
  \nn\\&& \mbox{}
         8\, \*  \colour4colour{\cas}  \*  \Big(
          \pgg(x) \* \Big[
            11/3\: \* \H(0)
          - 4 \* ( \Hh(0,0) + \Hh(1,0) + \H(2) )
            \Big]
          + [ 6 \* (1-x) - 22/3\, \* (x^{-1} - x^2) ]\, \* \H(0)
  \nn\\[-0.5mm]&& \mbox{}
          - 8 \* (1+x)\, \* \Hh(0,0)
        \Big)
      \:\: - \:\: 16/3\: \*  \colour4colour{\ca \* \nf}\,  \*
          \pgg(x)\, \* \H(0)
      \:\: + \:\: 8\, \*  \colour4colour{\cf \* \nf} \*  \Big(
            20/9\: \* x^{-1} + 3 + x - 56/9\: \* x^2
  \nn\\[-0.5mm]&& \mbox{}
          + [4 + 6\, \* x + 4/3\, \* (x^{-1} + x^2)]\, \* \H(0)
          + 2 \* (1 + x)\, \* \Hh(0,0)
        \Big)
\label{eq:dPgg1}
\eea
where we have used the abbreviation
$$
  p_{\rm gg}(x) \:\: = \:\: 1/(1-x) \, + \, 1/x \, - \, 2 
                            \, + \, x \, - \, x^2 \:\: .
$$
Note that the non-HPL terms in Eqs.~(\ref{eq:dPps1}) and (\ref{eq:dPgg1}) 
are identical up to an overall sign, cf.~Ref.~\cite{Dokshitzer:2005bf}. 
The functions $P_{\rm ps}^{\,T\!,\,S}$ denote the `pure singlet' contributions 
from which the quark-quark entries in, e.g., Eq.~(\ref{eq:Dsgevol}) are 
obtained by adding the corresponding non-singlet quantities.

The difference between the timelike NNLO pure-singlet splitting function and 
its spacelike counterpart of Ref.~\cite{Vogt:2004mw} reads
\bea
 \lefteqn{ \delta\, P^{\,(2)}_{\rm ps}(x) \;\; \equiv \;\;
           P^{\,(2)T}_{\rm ps}(x) -  P^{\,(2)S}_{\rm ps}(x) \;\; = \;\; }
  \nn\\&& \mbox{\hspn}
       + 8\, \*  \colour4colour{\ca \* \cf \* \nf }  \*  \Big(
            269/6\: \* x^{-1} + 14 + 113/2\, \* x - 346/3\: \* x^2
          + \z2\, \* ( 172 + 167\, \* x + 8\, \* x^2 )/3
  \nn\\[-0.5mm]&& \mbox{}
          - \z3\, \* (12\, \* x^{-1} - 13 + 65\, \* x - 28\, \* x^2)
        - 2 \* (1 + x)  \* \Big[
            16\: \* \zs2 
          + 4\, \* \Hhh(-1,0,0) 
          + 9\, \* \Hh(3,0) 
          + 4\, \* \Hh(3,1) 
  \nn\\[-0.5mm]&& \mbox{}
          + 10\, \* \z2 \* \H(2) 
          - 12\, \* \Hhh(2,0,0) 
          - 2\, \* \Hhh(2,1,0) 
          - 6\, \* \Hh(2,2) 
          - \H(4) 
          \Big]
        + 8/3\, \* (x^{-1} + x^2) \* \Big[
            4\, \* \Hhh(-1,0,0) 
          + \z2 \* \H(0)
          \Big]
  \nn\\[-0.5mm]&& \mbox{}
        - 2 \* (1 - x) \* \Big[
            8\, \* ( \Hh(-3,0) 
          + \Hhh(-2,0,0) )
          + 5\, \* \z2 \* \H(1) 
          - 9\, \* \z2 \* \H(0) 
          + 25/12\: \* \Hh(1,0) 
          - 6\, \* \Hhh(1,0,0) 
          - \Hhh(1,1,0) 
  \nn\\[-0.5mm]&& \mbox{}
          - 3\, \* \Hh(1,2) 
          + \Hh(2,1) 
          \Big]
        + 8/3 \* (x^{-1} - x^2) \* \Big[
            6\, \* \Hhh(1,0,0) 
          + \Hhh(1,1,0) 
          + 3\, \* \Hh(1,2) 
          - 5\, \* \z2 \* \H(1) 
          - \Hh(2,0)
          - \Hh(2,1) 
          \Big]
  \nn\\&& \mbox{}
          + 2/3\, \* (4\, \* x^{-1} + 27 - 63\, \* x 
            + 28\, \* x^2)\, \* \Hh(-2,0)
          - 2/3\, \* ( 20\, \* x^{-1} 
            - 27 + 9\, \* x + 56\, \* x^2)\, \* \Hh(-1,0)
  \nn\\[0.5mm]&& \mbox{}
          + (89/9\: \* x^{-1} + 55 + 1021/6\, \* x + 2297/18\: \* x^2
            - 46\, \* \z3 - 22\, \* x \* \z3)\, \* \H(0)
          - (8/9\: \* x^{-1}  
  \nn\\[0.5mm]&& \mbox{}
            + 293/6 + 370/3\, \* x + 538/9\: \* x^2
            + 2\, \* \z2 \* (1 - 7\, \* x) )\, \* \Hh(0,0)
          - 32\, \* x\, \* \H(0,0,0,0)
          + ( 5 - 16/3\: \* x^{-1}
  \nn\\[0.5mm]&& \mbox{}
            + 85\, \* x)\, \* \Hhh(0,0,0)
          - 1/6\, \* (115\, \* x^{-1} + 362 - 292\, \* x 
            - 185\, \* x^2)\, \* \H(1) 
          - (6\, \* x^{-1} + 48 + 59\, \* x
  \nn\\&& \mbox{}
            + 22\, \* x^2)\, \* \H(2) 
          - 2\, \* (5 + x - 8/3\, \* x^2)\, \* \Hh(2,0)
          + 4\,\* (2/3\: \* x^{-1} + x + 2\,\* x^2)\, \* \H(3)
          \Big)
  \nn\\&& \mbox{\hspn}
       + 8\, \*  \colour4colour{\cfs \* \nf }  \*  \Big(
          - 217/18 - 55/3\: \* x^{-1} + 122/9\, \* x + 101/6\: \* x^2
          + \z3\, \* (16\, \* x^{-1} + 36 + 24\, \* x)
  \nn\\[-0.5mm]&& \mbox{}
          - \z2\, \* (127 + 188\, \* x + 128\, \* x^2)/3
        + 2 \* (1 + x)  \* \Big[
            16\, \* \zs2 
          + 17\, \* \z3 \* \H(0) 
          + 8\, \* \z2\, \* x \* \H(0) 
          - 7\, \* \z2 \* \Hh(0,0) 
  \nn\\[-0.5mm]&& \mbox{}
          + 10\, \* \z2 \* \H(2) 
          + 9\, \* \H(2,0)
          - 12\, \* \Hhh(2,0,0) 
          - 2\, \* \Hhh(2,1,0) 
          - 6\, \* \Hh(2,2) 
          + 9\, \* \Hh(3,0) 
          + 4\, \* \Hh(3,1) 
          - \H(4) 
          \Big]
  \nn\\[-0.5mm]&& \mbox{}
        + 2 \* (1 - x)  \* 
            \Big[ 5\, \* \z2 \* \H(1) 
          + 2\, \* \H(2,0)
          + 139/12\: \* \Hh(1,0) 
          - 6\, \* \Hhh(1,0,0) 
          - \Hhh(1,1,0) 
          - 3\, \* \Hh(1,2) 
          + \Hh(2,1) 
          \Big]
  \nn\\[-0.5mm]&& \mbox{}
        + 8/3\, \* (x^{-1} - x^2) \* 
          \Big[ 5\, \* \z2 \* \H(1) 
          + 5/3\, \* \Hh(1,0) 
          - 6\, \* \Hhh(1,0,0) 
          - \Hhh(1,1,0) 
          - 3\, \* \Hh(1,2) 
          + 2\, \* \Hh(2,0) 
          + \Hh(2,1) 
          \Big]
  \nn\\&& \mbox{}
          - ( 527 + 2473\, \* x 
          + 811\, \* x^2 + 72\, \* \z2 )/18\: \* \H(0)
          + (62 + 81/2\, \* x + 208/9\: \* x^2 )\, \* \Hh(0,0)
  \nn\\[0.5mm]&& \mbox{}
          + (6 + 18\, \* x - 8\* x^2 )\, \* \H(0,0,0)
          + (385/18\: \* x^{-1} + 190/3 - 143/3\, \* x - 667/18\:  \* x^2)\,
             \* \H(1)
  \nn\\&& \mbox{}
          + (28/9\: \* x^{-1} + 71 + 46\, \* x 
            + 248/9\: \* x^2)\, \* \H(2)
          - 4/3 \* (4\: \* x^{-1} - 6 - 3\, \* x + 8\, \* x^2) \* \H(3)
          \Big) 
  \nn\\&& \mbox{\hspn}
       + 8\, \*  \colour4colour{\cf \* \nfs }  \*  \Big( 
            2/9\, \* (23\, \* x - 2\, \* x^{-1} - 20 - x^2)
        + 2 \* (1 + x) \* \Big[
            \z3 
          - \z2 \* \H(0) 
          - \H(1)
          + \Hh(2,0) 
          + \H(3) 
  \nn\\[-0.5mm]&& \mbox{}
          - \Hhh(0,0,0) 
          \Big]
          - (1 - x) \* ( \H(1) - \Hh(1,0) ) 
          + 4/3\, \*  (x^{-1} - x^2)\, \* \Hh(1,0)
          + 2/9\, \* (3 + 18\, \* x + 10\, \* x^2)\, \* \H(0)
  \nn\\[-0.5mm]&& \mbox{}
          - ( 7 + x - 4\, \* x^2)/3\: \* \Hh(0,0)
          - (20\, \* x^{-1} - 56\, \* x^2)/9\: \* \H(1)
          + (3 + 7\, \* x + 8/3\: \* x^2)\, \* (\z2 - \H(2))
          \Big)
\:\: .\label{eq:dPps2}
\eea
The corresponding result for the gluon-gluon splitting functions is given by
\bea
 \lefteqn{ \delta\, P^{\,(2)}_{\rm gg}(x) \;\; \equiv \;\;
           P^{\,(2)T}_{\rm gg}(x) -  P^{\,(2)S}_{\rm gg}(x) \;\; = \;\; }
  \nn\\&& \mbox{\hspn}
       + 16\, \*  \colour4colour{\cat}  \*  \Big(
          \pgg(x) \* \Big[
          ( 1025/54 - 11/3\, \* \z2 - 2 \* \z3 )\, \* \H(0)
          - 49/3\: \* \Hh(0,0)
          - 33\, \* \Hhh(0,0,0)
          + 16\, \* \Hhhh(0,0,0,0)
  \nn\\&& \mbox{}
          - ( 268/9 - 8 \* \z2) \, \* ( \Hh(1,0) + \,\H(2) )
          - 44/3\: \* ( \Hhh(1,0,0) + \Hh(2,0) + \H(3) )
          + 12\, \* \Hhhh(1,0,0,0)
          + 4\, \* \Hhh(2,0,0)
  \nn\\&& \mbox{}
          + 4\, \* \Hh(3,0)
          + 12\, \* \H(4)
          \Big]
        + \pgg(-x) \* \Big[
            16\, \* \Hh(-3,0)
          - 16\, \* \z2 \* \H(-2)
          - 16\, \* \Hhh(-2,-1,0)
          - 22/3\: \* \Hh(-2,0)
  \nn\\&& \mbox{}
          + 28\, \* \Hhh(-2,0,0)
          + 8\, \* \Hh(-2,2)
          - 16\, \* \Hhh(-1,-2,0)
          - 32\, \* \Hhhh(-1,-1,0,0)
          - 16\, \* \z2 \* \Hh(-1,0)
          - 44/3\: \* \Hhh(-1,0,0)
  \nn\\[0.5mm]&& \mbox{}
          + 36\, \* \Hhhh(-1,0,0,0)
          + 8\, \* \Hhh(-1,2,0)
          + 16\, \* \Hh(-1,3)
          + (14 \* \z3 - 11/3\, \* \z2)\, \* \H(0)
          + 16 \* \z2\, \* \Hh(0,0)
          + 11\, \* \Hhh(0,0,0)
  \nn\\&& \mbox{}
          - 16\, \* \Hhhh(0,0,0,0)
          - 4\, \* \Hh(3,0)
          - 12\, \* \H(4)
          \Big]
        + (1 + x) \* \Big[
          - 24\,\* \Hh(-2,0) 
          - 48\, \* \Hhh(-1,0,0) 
          + 14/3\: \* \Hh(2,0) 
  \nn\\[-0.5mm]&& \mbox{}
          + 28/3\: \* \H(3) 
          \Big]
        + (1 - x) \* \Big[
            32\, \* ( \Hh(-3,0) + \Hhh(-2,0,0) )
          - ( 881/36 - 24 \* \z3 )\, \* \H(0)
          - 27 \* ( \Hh(1,0) + \H(2) )
          \Big]
  \nn\\[-0.5mm]&& \mbox{}
        - 44/3\: \* (x^{-1} + x^2) \* \Big[
            2\, \* \Hh(-2,0)
          + 4\, \* \Hhh(-1,0,0)
          - \Hh(2,0)
          - 2\, \* \H(3)
          \Big]
        + (x^{-1} - x^2) \* \Big[
            2261/54\: \* \H(0)
  \nn\\[-0.5mm]&& \mbox{}
          + 134/9\, \* ( \Hh(1,0) + \H(2) )
          \Big]
          - (44\, \* x^{-1} + 86 + 14\, \* x 
            + 132\, \* x^2)/3\: \* \z2 \* \H(0)
          + (536\, \* x^{-1} + 425 
  \nn\\[-0.5mm]&& \mbox{}
          + 515\, \* x + 752\, \* x^2 + 288 \* \z2 )/9\: \* \Hh(0,0)
          + (88\, \* x^{-1} - 10 + 8\, \* x + 44\, \* x^2)\, \* \Hhh(0,0,0) 
          + 64 \* x\, \* \Hhhh(0,0,0,0) \Big)
  \nn\\&& \mbox{\hspn}
       + 16\, \*  \colour4colour{\cas \* \nf}  \*  \Big(
          \pgg(x) \* \Big[ 
          - ( 158/27 - 2/3\, \* \z2 )\, \* \H(0)
          - 4/9\: \* \Hh(0,0)
          + 6\, \* \Hhh(0,0,0)
          + 40/9\, \* ( \Hh(1,0) + \H(2) )
  \nn\\[-0.5mm]&& \mbox{}
          + 8/3\, \* ( \Hhh(1,0,0) + \Hh(2,0) + \H(3) )
          \Big]
       + 2/3\, \* \pgg(-x) \* \Big[
            2\, \* \Hh(-2,0)
          + 4\, \* \Hhh(-1,0,0)
          + \z2 \* \H(0)
          - 3\, \* \Hhh(0,0,0)
          \Big]
  \nn\\[-0.5mm]&& \mbox{}
       - {4 \over 3}\, \* (1 + x) \* \Big[
            \z2 \* \H(0) 
          - \Hh(2,0) 
          - 2\, \* \H(3) 
          \Big]
       - (1 - x) \* \Big[ 
            173/9\: \* \H(0) 
          + 2\, \* ( \Hh(1,0) 
          + \H(2) )
          \Big]
        + (x^{-1} - x^2) \* 
  \nn\\[-0.5mm]&& \mbox{}
          \: \Big[
          913/54\: \* \H(0)
          + 26/9\, \* ( \Hh(1,0) + \H(2) )
          \Big]
          + 4/9\, \* ( 35\, \* x^{-1} + 21 + 48\, \* x )\, \* \Hh(0,0)
          + 4\, \* ( 1 + 4 \* x)\, \* \Hhh(0,0,0)
          \Big)
  \nn\\&& \mbox{\hspn}
       \: + \: 
       {16 \over 27}\: \*  \colour4colour{\ca \* \nfs}  \*  \Big(
          \pgg(x) \* \Big[
            10\, \* \H(0) 
          + 12\, \* \Hh(0,0)
          \Big] 
          + 12 \* (1 + x)\, \* \Hh(0,0)
          + ( 13\, \* (x^{-1} - x^2) 
          - 9 + 9\, \* x) \, \* \H(0)
          \Big)
  \nn\\&& \mbox{\hspn}
       + 8\, \*  \colour4colour{\ca \* \cf \* \nf}  \*  \Big(
         - 2 \* \pgg(x)\: \* \H(0)
          - 269/6\: \* x^{-1} - 14 - 113/2\, \* x + 346/3\: \* x^2
          - \z2\, \* ( 172 + 167\, \* x 
  \nn\\[-0.5mm]&& \mbox{}
            + 8\, \* x^2 )/3
          + \z3\, \* (12\, \* x^{-1} - 13 + 65\, \* x - 28\, \* x^2)
        + 2 \* (1 + x)  \* \Big[
            16\, \* \zs2
          + 2\, \* \Hh(-2,0) 
          - 4\, \* \Hhh(-1,0,0) 
  \nn\\&& \mbox{}
          + 17\, \* \z3 \* \H(0)
          + 4/3\: \* \z2 \* \H(0)
          - 3\, \* \z2 \* \Hh(0,0)
          + 10\, \* \z2 \* \H(2) 
          - 12\, \* \Hhh(2,0,0) 
          - 2\, \* \Hhh(2,1,0) 
          - 6\, \* \Hh(2,2) 
          + 4\, \* \Hh(3,1) 
  \nn\\&& \mbox{}
          + 9\, \* \Hh(3,0) 
          - \H(4) 
          \Big]
        + 8/3\, \* (x^{-1} + x^2 ) \* \Big[
            4\, \* \H(-1,0,0)
          + \z2 \* \H(0)
          \Big]
        - 2 \* (1 - x) \* \Big[
            8 \* ( \Hh(-3,0) + \Hhh(-2,0,0) )
  \nn\\[-0.5mm]&& \mbox{}
          + 18\, \* \Hh(-2,0) 
          + 9\, \* \z2 \* \H(0)
          + 6\, \* \z3 \* \H(0)
          + 4\, \* \z2 \* \H(0,0)
          - 145/12\: \* \Hh(1,0) 
          \Big]
       + \Big[ \, { 8 \over 3} \* (x^{-1} - x^2 ) + 2 \* (1 - x) \Big] 
  \nn\\[-0.5mm]&& \mbox{}
            \* \: \Big[
            3\, \* \Hh(-2,0)
          - 11/3\: \* \Hh(1,0) 
          + 5\, \* \z2 \* \H(1)
          - 6\, \* \Hhh(1,0,0) 
          - \Hhh(1,1,0) 
          - 3\, \* \Hh(1,2) 
          + \Hh(2,1)
          \Big]
          + ( 40\,\* x^{-1} - 54 
  \nn\\&& \mbox{}
            + 18\, \* x + 112\: \* x^2 )/3 \: \* \Hh(-1,0) 
          - ( 59\, \* x^{-1} + 45 + 1081/6\, \* x 
            + 157/2\: \* x^2 )\, \* \H(0) 
          - (464\, \* x^{-1} 
  \nn\\[0.5mm]&& \mbox{}
            + 329/2 - 146\, \* x - 66\, \* x^2 )/9\: \* \Hh(0,0)
          - (80/3\: \* x^{-1} - 17 + 15\, \* x )\,\* \Hh(0,0,0)
          - 32\, \* x\, \* \Hhhh(0,0,0,0) 
  \nn\\[0.5mm]&& \mbox{}
          + (115\, \* x^{-1} + 362 - 292\, \* x - 185\,\* x^2 )/6\: \* \H(1)
          - 1/9\: \* (34\, \* x^{-1} - 546 - 417\, \* x
            - 286\,\* x^2 )\, \* \H(2)
  \nn\\&& \mbox{}
          + (8\, \* x^{-1} + 10 - 14\, \* x - 24\, \* x^2 )/3\: \* \Hh(2,0)
          - 8/3\, \* (x^{-1} + 5 + 13/2\, \* x + 3\, \* x^2)\, \* \H(3)
        \Big)
  \nn\\&& \mbox{\hspn}
       + 8\, \*  \colour4colour{\cfs \* \nf }  \*  \Big(
            217/18 + 55/3\: \* x^{-1} - 122/9\, \* x - 101/6\: \* x^2
          - \z3\, \* (16\, \* x^{-1} + 36 + 24\, \* x)
          + \z2 / 3 \* 
  \nn\\[-0.5mm]&& \mbox{}
          \:  (127 + 188\, \* x + 128\, \* x^2)
        - 2 \* (1 + x)  \* \Big[
            16\, \* \zs2
          + \z2 \* \Hh(0,0) 
          + 10\, \* \z2 \* \H(2) 
          + 17\, \* \z3 \* \H(0) 
          - 12\, \* \Hhh(2,0,0) 
  \nn\\[-0.5mm]&& \mbox{}
          - 2\, \* \Hhh(2,1,0) 
          - 6\, \* \Hh(2,2) 
          + 9\, \* \Hh(3,0) 
          + 4\, \* \Hh(3,1) 
          - \H(4) 
          \Big]
       - \Big[ \, { 8 \over 3} \* (x^{-1} - x^2 ) + 2 \* (1 - x) \Big] \* \Big[ 
            5\, \* \z2 \* \H(1) 
  \nn\\[-0.5mm]&& \mbox{}
          + 3\, \* \Hh(1,0)
          - 6\, \* \Hhh(1,0,0)
          - 3\, \* \Hh(1,2)
          - \Hhh(1,1,0)
          + \Hh(2,1)
          \Big]
          + (4\, \* x^{-1} + 283/6 + 239/2\: \* x + 739/18\: \* x^2
  \nn\\&& \mbox{}
          - 8\, \* \z2 - 20\, \* \z2\, \* x - 16/3\, \* \z2\, \* x^2)\,\* \H(0)
          - ( 18 + 97/2\: \* x + 16\, \* x^2 )\, \* \Hh(0,0) 
          - ( 6 - 6\, \* x + 8\, \* x^2 )\, \* \Hhh(0,0,0) 
  \nn\\[0.5mm]&& \mbox{}
          - (385\: \* x^{-1} + 1140 - 858\, \* x - 667\, \* x^2 )/18\: \* \H(1)
          + 53/6\: \* (1 - x) \* \Hh(1,0)
          - ( 20/3\: \* x^{-1} + 45 
  \nn\\&& \mbox{}
             + 72\, \* x + 24\, \* x^2 )\, \* \H(2)
          - (32/3\: \* x^{-1} + 14 + 6\, \* x )\, \* \Hh(2,0) 
          - (16/3\: \* x^{-1} - 8 - 12\, \* x )\, \* \H(3) 
        \Big)
  \nn\\&& \mbox{\hspn}
       + 8/9\: \*  \colour4colour{\cf \* \nfs }  \*  \Big(
            4\, \* x^{-1} + 40 - 46\, \* x + 2\, \* x^2
          - 9 \* \z2\, \* (3 + 7\, \* x + 8/3\: \* x^2)
       - 6 \* (1 + x) \* \Big[
            3\, \* \z3
          + \z2 \* \H(0)
  \nn\\[-0.5mm]&& \mbox{}
          + 3\, \* \Hhh(0,0,0)
          - \H(2,0)
          - 5\, \* \H(3)
          \Big]
          - ( 92/3\: \* x^{-1} - 6 + 48\, \* x - 32/3\, \* x^2 )\, \* \H(0)
          - ( 16\, \* x^{-1} + 83 
  \nn\\&& \mbox{}
            + 101\, \* x + 28\, \* x^2 )\, \* \Hh(0,0) 
          + ( 20\, \*  x^{-1} + 27 + 9\, \* x - 56\, \* x^2 )\, \* \H(1)
          + ( 4\, \* x^{-1} + 3 - 3\, \* x - 4\, \* x^2 )\, \* \Hh(1,0) 
  \nn\\&& \mbox{}
          + ( 16\, \* x^{-1} + 39 + 51\, \* x + 8\, \* x^2 )\, \* \H(2) 
        \Big)
\:\: .\label{eq:dPgg2}
\eea
Eqs.~(\ref{eq:dPps2}) and (\ref{eq:dPgg2}) represent the main new results of 
this letter. Analogous to the non-singlet result of Ref.~\cite{Mitov:2006ic}, 
the coefficient of $ \cat\, p_{\rm gg}(x) H_{0,0}\,\z2$ in Eq.~(\ref{eq:dPgg2}) 
differs from the result of the analytic continuation as specified above.
We have determined the correct coefficient via the momentum sum rule for 
$\nf = 0$. The complete $\cat$, $\cas \nf$ and $\ca \nfs$ parts of 
Eq.~(\ref{eq:dPgg2}) have been independently derived by applying the approach 
of Ref.~\cite{Dokshitzer:2005bf} to the (non-singlet like) non-$\cf$ 
contributions to $P_{\rm gg}$. 
Furthermore an inspection of the above results reveals that also $\delta\, 
P^{\,(2)}_ {\rm gg} + \delta\, P^{\,(2)}_{\rm ps}$ does not receive any non-HPL 
contributions. Since the determination of these two functions uses largely 
independent information, we view this fact as another non-trivial check of our 
calculations.

As predicted in Ref.~\cite{Dokshitzer:2005bf}, the large-$x$ behaviour of 
$P^{\,(2)T}_ {\rm gg}$ is identical to that of its spacelike counterpart 
(see Eq.~(4.16) of Ref.~\cite{Vogt:2004mw}) up to the sign of the subleading 
$\ln (1-x)$ contribution. The pure-singlet splitting functions are suppressed 
by two powers of $\,(1-x)\,$ for $\,x\ra 1$. 
We  now turn to the small-$x$ limits. Up to terms suppressed by powers of $x$
the NLO quantities read
\bea
 xP^{\,(1)T}_{\rm qq}(x) & =\! & \mbox{} 
  - {80 \over 9}\: \* \cf \* \nf
\:\: , \nn \\
 xP^{\,(1)T}_{\rm gg}(x) & =\! & \mbox{} 
  - 16\, \* \cas\: \* L_0^2
  \: - \: {8 \over 3} \* 
     \Big[ 11\: \* \cas + 2\, \* \nf\, \* (\ca - 2\, \* \cf ) \Big]\, \* L_0
  \: - \: {92 \over 9}\: \* \nf\, \* ( \ca - 2\, \* \cf )
\quad
\eea
with $L_0 \equiv \ln x$. 
The log-enhanced contributions of our new diagonal NNLO splitting functions are
\bea
 xP^{\,(2)T}_{\rm qq}(x) & = &
   - {32 \over 9}\: \* \ca \* \cf \* \nf\, \* ( 2\, \* L_0^3 + L_0^2 )
   \: + \: {8 \over 27}\: \* (155 + 72\, \* \z2)\, \* \ca \* \cf \* \nf\: \* L_0
\: + \: {\cal O}(1) 
\:\: ,
\\[2mm]
 xP^{\,(2)T}_{\rm gg}(x) & = &
     {64 \over 3}\: \* \cat\: \* L_0^4 
   \: + \: {32 \over 9}\: \* ( 33\, \* \cat + 6\, \* \cas \* \nf 
      - 10\, \* \ca \* \cf \* \nf )\: \* L_0^3
 \nn \\ & & \mbox{\hspn}
   + {8 \over 9}\: \* \Big[ ( 389 - 144\, \* \z2 )\, \* \cat
     \: + \: 136\, \* \cas \* \nf 
     \: - \: 232\, \* \ca \* \cf \* \nf 
     \: + \: 4\, \* \nfs \* (\ca - 2\, \* \cf) \Big]\, \* L_0^2
 \nn \\ & & \mbox{\hspn}
   + {8 \over 27}\: \* \Big[ (4076 - 990\, \* \z2 - 972\, \* \z3)\, \* \cat
     \: + \: ( 739 - 36\, \* \z2 )\, \* \cas \* \nf
 \nn \\ & & \mbox{}
     \: - \: ( 1819 - 144\, \* \z2 )\, \* \ca \* \cf \* \nf
     \: + \: 108\, \* \cfs \* \nf
     \: + \: 46\, \* \nfs \* (\ca - 2\, \* \cf)  \Big]\, \* L_0 
\: + \: {\cal O}(1) 
\label{eq:Pgg2x0} \:\: . \quad
\eea 
Thus, in contrast to the corresponding spacelike quantities, the NNLO timelike 
singlet splitting functions receive double-logarithmic contributions with very 
large coefficients, with the leading term of Eq.~(\ref{eq:Pgg2x0}) agreeing 
with Refs.~\cite{Mueller:1981ex,Bassetto:1982ma}.
Consequently both $\,xP^{\,(2)T}_{\rm ps}\!$ and $\,xP^{\,(2)T}_{\rm gg}\!$ 
-- despite a large cancellation between the leading- and subleading logarithms 
for the latter quantity -- show a huge enhancement already at $x \gsim 10^{-3}$
as illustrated in Fig.~\ref{pic:P2tsx0}. For this and the next figure the 
harmonic polylogarithms have been evaluated using the {\sc Fortran} package of 
Ref.~\cite{Gehrmann:2001pz}.
    
\begin{figure}[hbp]
\vspace*{-1mm}
\centerline{\epsfig{file=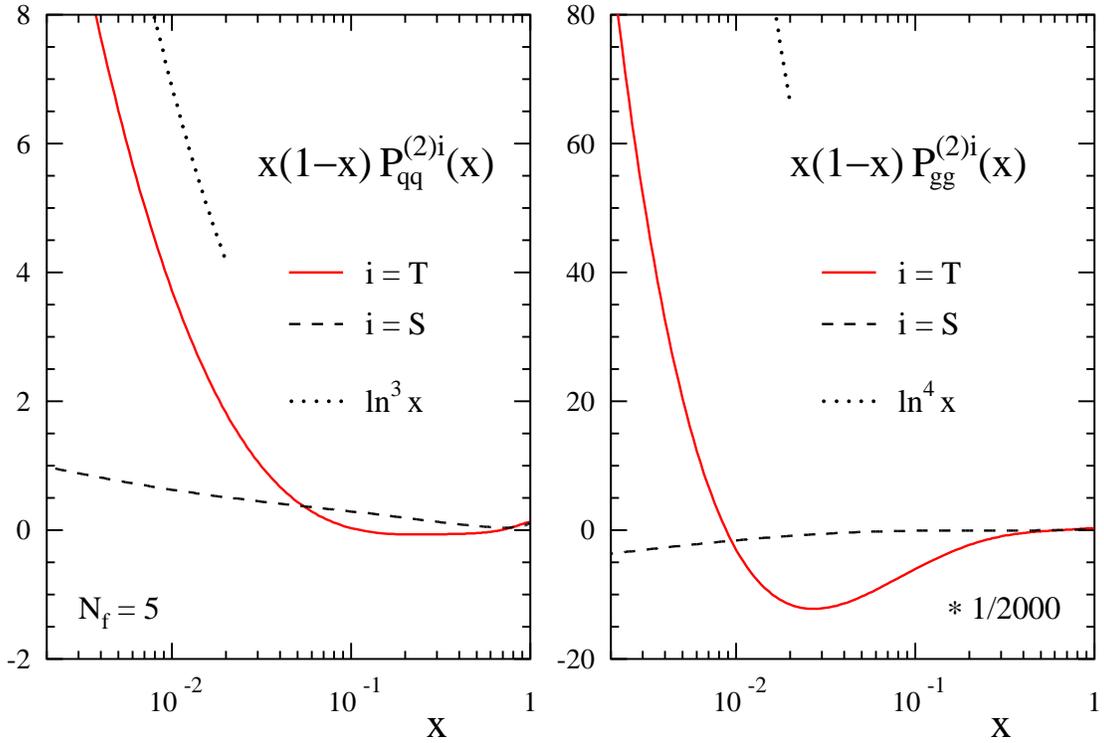,width=15.0cm,angle=0}}
\vspace{-3mm}
\caption{ \label{pic:P2tsx0}
 The third-order timelike quark-quark and gluon-gluon splitting functions 
 for five flavours, multiplied by $x\,(1-x)$ and divided by $2000 \simeq 
 (4\pi)^3$ for display purposes. Also shown are the respective leading 
 small-$x$ contributions and the corresponding spacelike splitting functions.}
\vspace{2mm}
\end{figure}
\begin{figure}[thp]
\vspace{-2mm}
\centerline{\epsfig{file=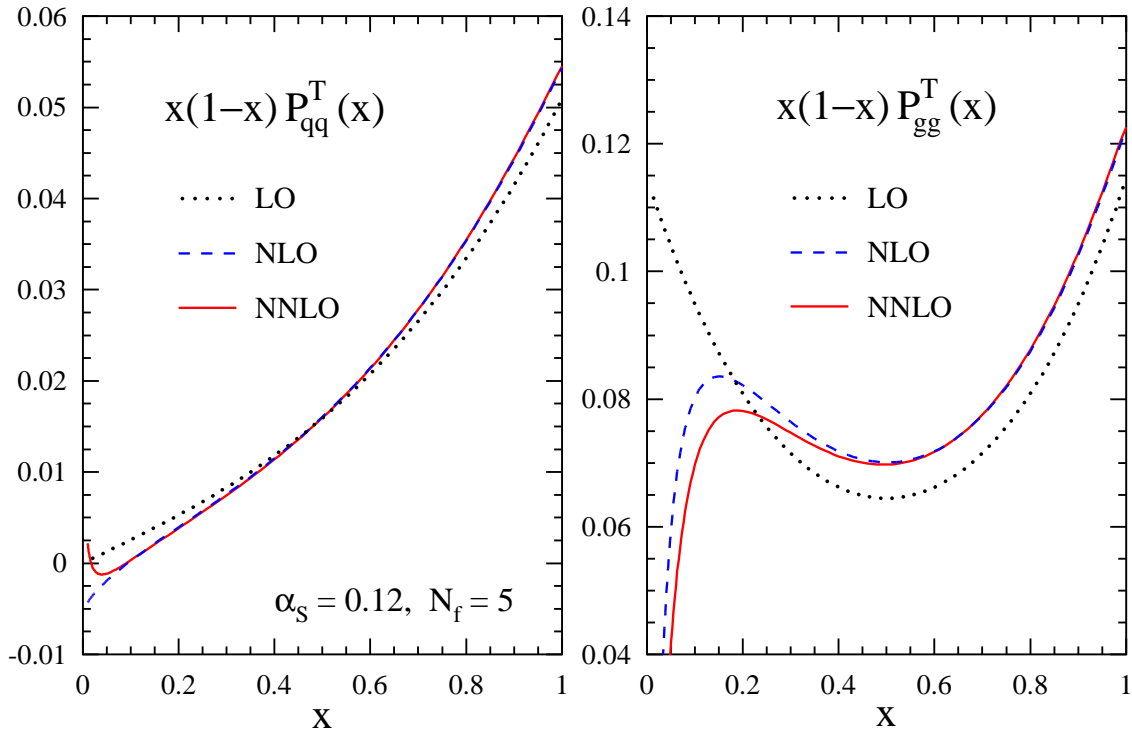,width=15.0cm,angle=0}}
\vspace{-3mm}
\caption{ \label{pic:Ptexp}
 The perturbative expansion of the timelike quark-quark and gluon-gluon 
 splitting functions, again multiplied by $x\,(1-x)$, at a typical value of 
 the strong coupling constant.}
\vspace{-2mm}
\end{figure}

Returning to the region of medium and large values of $x$, Fig.~\ref{pic:Ptexp}
shows the LO, NLO and NNLO approximations to the diagonal entries in Eq.~(\ref
{eq:Dsgevol}) at a scale relevant to gauge-boson and, maybe, Higgs decay. 
Also here the higher-order corrections are larger than for the spacelike case, 
in particular for the gluon-gluon splitting. Nevertheless the perturbative 
expansion appears well-behaved at least for $\,x \gsim 0.1$. As usual, the 
region of safe applicability of the NNLO approximation for Eq.~(\ref{eq:Devol}) 
will be wider as an effect of the Mellin convolution.

As mentioned above, we are presently not in a position to derive the 
$\z2$-terms of the off-diagonal quantities $P^{\,(2)T}_{\rm qg}(x)$ and 
$P^{\,(2)T}_{\rm gq}(x)$ (with the exception of some $\nf$-enhanced 
contributions). However, their non-$\z2$ second moments provide another check, 
via the momentum sum rule, of our new results (\ref{eq:dPps2}) and 
(\ref{eq:dPgg2}). Furthermore, using these results and the momentum sum rule, 
the missing off-diagonal terms can be reconstructed at $\,N = 2$. Hence we can 
write down the complete NNLO expressions for the second moments of all four 
splitting splitting functions, 
\bea
 \lefteqn{ 
  P^{\,(2)T}_{\rm qq}(N\! =\! 2) \:\: =\:\: - P^{\,(2)T}_{\rm gq}(N\! =\! 2)
  \;\; = \;\; \mbox{}  
       - \cft \: \* \Bigg(
            {54556 \over 243}
          - {7264 \over 27}\: \* \z2
          - 320\, \* \z3
          + 256\, \* \zs2
          \Bigg)
} \qquad
\nn \\ & & \mbox{\hspn}
        - \cfs\*\ca \: \* \Bigg(
            {6608 \over 243} 
          - {2432 \over 9}\: \* \z2
          + {2464 \over 9}\: \* \z3
          - {128 \over 3}\: \* \zs2
          \Bigg)
       \: - \: \cf\*\cas \: \* \Bigg(
            {20920 \over 243}
          + {64 \over 3}\: \* \z3
          \Bigg)
\nn \\ & & \mbox{\hspn}
       - \cf \* \ca \* \nf \: \* \Bigg(
            {55 \over 81}
          + {296 \over 27}\: \* \z2
          - {512 \over 9}\: \* \z3
          \Bigg)
       \: - \: \cfs \* \nf \: \* \Bigg(
            {2281 \over 81}
          - {32 \over 9}\: \* \z2 + {64 \over 9}\: \* \z3 
          \Bigg)
\:\: , \\[2mm]
 \lefteqn{
  P^{\,(2)T}_{\rm gg}(N\! =\! 2) \:\: =\:\: - P^{\,(2)T}_{\rm qg}(N\! =\! 2)
  \;\; = \;\; \mbox{}
       - \cas \* \nf\: \* \Bigg(
            {6232 \over 243}
          - {2132 \over 27}\: \* \z2
          - {128 \over 9}\: \* \z3
          + {160 \over 3}\: \* \zs2
          \Bigg) 
} \qquad
\nn \\ & & \mbox{\hspn}
       + \: \ca \* \nfs \: \* \Bigg(
            {2 \over 27}
          - {160 \over 27}\: \* \z2
          + {64 \over 9}\: \* \z3
          \Bigg)
       \: - \: \ca \* \cf \* \nf \: \* \Bigg(
            {2681 \over 243}
          - {760 \over 27}\: \* \z2
          + {56 \over 9}\: \* \z3
          \Bigg)
\nn \\ & & \mbox{\hspn}
       - \: \cfs \* \nf \: \* (
            {10570 \over 243}
          - {352 \over 27}\: \* \z2
          - {32 \over 9}\: \* \z3
          )
       \: - \: \cf \* \nfs \* \Bigg(
            {41 \over 9}
          - {128 \over 27}\: \* \z2
          \Bigg)
\:\: .
\eea
After inserting the numerical values of the QCD colour factors and the
Riemann $\zeta$-function, these results yield, for five light flavours, the 
benign perturbative expansion
\bea
 P^{\,T}_{\rm gq}(N\! = \! 2,\: \nf\! =\! 5) & \simeq & 
   8\als/(9\pi) \; \left( 1 \: - \: 0.687\, \als \: + \: 0.447\, \als^2 
  \: + \: \ldots \,\right)
\nn \\[0.5mm]
 P^{\,T}_{\rm qg}(N\! =\! 2,\: \nf\! =\! 5) & \simeq & 
   5\als/(6\pi) \; \left( 1 \: - \: 1.049\, \als \: + \: 1.163\, \als^2 
  \: + \: \ldots \,\right)
  \:\: . 
\eea

We finally turn to the timelike gluon coefficient functions in 
Eqs.~(\ref{eq:FphiT1}) -- (\ref{eq:FphiT3}) and their quark analogues 
$c^{\,(n)T}_{\phi,\rm q}\!(x)$. In the limit of a heavy top quark and 
negligible masses of all other flavours, the coefficient functions for the 
standard-model Higgs-boson differ from those of this scalar $\phi$ only by a 
perturbative prefactor known to N$^3$LO \cite{Chetyrkin:1997un}. Thus the
latter quantities are directly relevant to one-particle inclusive Higgs decay,
and the second-moment combination 
$$
  (\, C^{\, T}_{\phi,q} + C^{\, T}_{\phi,g\,} ) (N\!=\!2) \;\: = \;\:
  1 \: + \: \as \, c_{\phi}^{\,(1)}
    \: + \: \as^2 \, c_{\phi}^{\,(2)}
    \: + \: \as^3 \, c_{\phi}^{\,(3)}  \: + \: \ldots \:\: ,
$$
directly enters the Higgs decay rate in the above limit. For the expansion
coefficients $c_{\phi}^{\,(n)}$ our analytic continuations lead to 
\bea
\label{eq:cphi1} 
  c_{\phi}^{\,(1)} &\! =\! & 
       {73 \over 3}\: \* \ca 
     - {14 \over 3}\: \* \nf 
 \:\: , \\[2mm]
\label{eq:cphi2} 
  c_{\phi}^{\,(2)} &\! =\! &
         \cas \: \* \Bigg(
            {37631 \over 54}
          - {242 \over 3}\: \* \z2
          - 110\, \* \z3
          \Bigg)
       \: - \: \ca \* \nf \: \* \Bigg(
            {6665 \over 27}\,
          - {88 \over 3}\: \* \z2
          + 4\, \* \z3
          \Bigg)
\nn \\ & & \mbox{\hspn\hspn}
       - \: \cf \* \nf \: \* \Bigg(
            {131 \over 3}
          - 24\, \* \z3
          \Bigg)
       \: + \: \nfs \: \* \Bigg(
            {508 \over 27}
          - {8 \over 3}\: \* \z2
          \Bigg)
 \:\: ,\\[2mm]
\label{eq:cphi3} 
  c_{\phi}^{\,(3)} &\! =\! &
          f(\z2)  \: + \: 
         \cat \: \* \Bigg(
            {15420961 \over 729}
          - {178156 \over 27}\:\* \z3
          + {3080 \over 3}\: \* \z5
          \Bigg)
       + \: \ca \* \nfs \: \* \Bigg(
            {413308 \over 243}
          + {56 \over 9}\: \* \z3
          \Bigg)
\nn \\ & & \mbox{\hspn\hspn}
       \: - \: \cas \* \nf \: \* \Bigg(
            {2670508 \over 243}\:
          - {9772 \over 9}\: \* \z3
          + {80 \over 3}\: \* \z5
          \Bigg)
       \: - \: \cf \* \ca \* \nf\: \* \Bigg(
            {23221 \over 9}\:
          - 1364\: \* \z3
          - 160\: \* \z5
          \Bigg)
\nn \\ & & \mbox{\hspn\hspn}
       \: + \: \cfs \* \nf \: \* \Bigg(
            {221 \over 3}\:
          - 320\: \* \z5
          + 192\: \* \z3
          \Bigg)
       + \: \cf \* \nfs \: \* (
            440\:
          - 240\: \* \z3
          )
       \: - \: \nft \: \* \Bigg(
            {57016 \over 729}
          - {64 \over 27}\: \* \z3
          \Bigg)
\qquad
\eea
with a function $f(\z2)$ which we cannot derive at this point. 
Eq.~(\ref{eq:cphi1}) reproduces a well-known NLO result of Refs.~\cite
{Inami:1982xt,Djouadi:1991tka}. Our coefficient (\ref{eq:cphi2}) represent a
completely independent re-derivation of the NNLO expression first obtained
in Ref.~\cite{Chetyrkin:1997iv} (see also Ref.~\cite{Schreck:2007um}).
Finally the third-order result (\ref{eq:cphi3}) provides, despite the missing 
$\z2$-contributions, a highly non-trivial confirmation of the recent N$^3$LO 
calculation of Ref.~\cite{Baikov:2006ch} 
to which the reader is referred for a further discussion.

To summarize, we have used relations between spacelike and timelike quantities
to compute some important higher-order QCD corrections for timelike processes. 
We expect that further progress can be achieved along these lines. 
However this will require improvements on the present decompositions into 
purely real and real-virtual terms which are beyond the scope of this letter.

{\sc Form} and {\sc Fortran} files of our results can be obtained from 
$\,${\tt http://arXiv.org}$\,$ by downloading the source of this article. 
Furthermore they are available from the authors upon request.

\vspace{2mm}
\noindent
{\bf Acknowledgements:}~
This research draws upon (partly still unpublished) results obtained 
together with J.~Vermaseren whom we would like to thank for a very pleasant 
collaboration. We also thank A. Mitov for stimulating discussions.
The work of S.M. has been supported by the Helm\-holtz Gemeinschaft (contract 
VH-NG-105) and partly by the Deutsche Forschungsgemeinschaft (SFB/TR~9). 
During part of this research A.V. enjoyed the hospitality of the 
Instituut-Lorentz of Leiden University and S.M. that of the Galileo-Galilei
Institute in Florence.

\end{document}